\documentclass[]{spie}  %>>> use for US letter paper
\usepackage{amsmath,amsfonts,amssymb}
\usepackage{graphicx}
\usepackage[colorlinks=true, allcolors=blue]{hyperref}

\title{The path toward 500~um depletion of AstroPix, a pixelated silicon HVCMOS sensor for space and EIC}

\author[a,b]{Amanda L. Steinhebel}
\author[c]{Jennifer Ott}
\author[c]{Olivia Kroger}
\author[b]{Regina Caputo}
\author[c]{Vitaliy Fadeyev}
\author[c]{Anthony Affolder}
\author[c]{Kirsten Affolder}
\author[c]{Aware Deshmukh}
\author[d]{Nicolas Striebig}
\author[e]{Manoj Jadhav}
\author[f]{Yusuke Suda}
\author[f]{Yasushi Fukazawa}
\author[e]{Jessica Metcalfe}
\author[d]{Richard Leys}
\author[d]{Ivan Peric}
\author[c]{Taylor (K.-W.) Shin}
\author[a,b]{Daniel Violette}

\affil[a]{NASA Postdoctoral Program Fellow}
\affil[b]{NASA Goddard Space Flight Center, Greenbelt, MD, USA}
\affil[c]{Santa Cruz Institute for Particle Physics (SCIPP), University of California Santa Cruz, Santa Cruz, CA, USA}
\affil[d]{Karlsruhe Institute of Technology, Karlsruhe, Germany}
\affil[e]{Argonne National Laboratory, Lemont, IL, USA}
\affil[f]{Physics Program, Graduate School of Advanced Science and Engineering, Hiroshima University, 739-8526 Hiroshima, Japan}

\authorinfo{Further author information: (Send correspondence to A.L.S.)\\A.L.S.: E-mail: amanda.l.steinhebel@nasa.gov\\  R.C.: E-mail: regina.caputo@nasa.gov}

\begin{document} 
\maketitle

%%%%%%%%%%%%%%%%%%%%%%%%%%%%%%%%%%%%%%%%%%%%%%%%%%%%%%%%%%%%%%%%%%%%%%%%%%%%%%%%%%%%%%%%%%%%%%%%%%%%%%%%%%%%%%%%%%%%%%%%%%%%%%%%%%%%%%%%%%%%%%%%%%%%%%%%%%%%%%%%%%%%%%%%%%%%%%%%%%%%%%%%%%%%%%%%%%%%%%%%%%%%%%%%%%%%%%%%%%%%%%%%%%%%%%%%%%%%%%%%%%%%%%%%%%%%%%%%%%%%%%%%%%%%%%%%%
\begin{abstract}
The precise reconstruction of Compton-scatter events is paramount for an imaging medium-energy gamma-ray telescope. The proposed AMEGO-X is enabled by a silicon tracker utilizing AstroPix chips - a pixelated silicon HVCMOS sensor novel for space use. To achieve science goals, each $500 \times 500~\mu\mathrm{m}^2$ pixel must be sensitive for energy deposits ranging from 25 - 700 keV with an energy resolution of 5 keV at 122 keV ($<10\%$). This is achieved through depletion of the 500 $\mu$m thick sensor, although complete depletion poses an engineering and design challenge. This talk will summarize the current status of depletion measurements highlighting direct measurement with TCT laser scanning and the agreement with simulation. Future plans for further testing will also be identified.  
\end{abstract}

% Include a list of keywords after the abstract 

%%%%%%%%%%%%%%%%%%%%%%%%%%%%%%%%%%%%%%%%%%%%%%%%%%%%%%%%%%%%%%%%%%%%%%%%%%%%%%%%%%%%%%%%%%%%%%%%%%%%%%%%%%%%%%%%%%%%%%%%%%%%%%%%%%%%%%%%%%%%%%%%%%%%%%%%%%%%%%%%%%%%%%%%%%%%%%%%%%%%%%%%%%%%%%%%%%%%%%%%%%%%%%%%%%%%%%%%%%%%%%%%%%%%%%%%%%%%%%%%%%%%%%%%%%%%%%%%%%%%%%%%%%%%%%%%%
\section{INTRODUCTION}
\label{sec:intro} 

\begin{figure} [ht]
   \begin{center}
   \begin{tabular}{ccc} %% tabular useful for creating an array of images 
   \includegraphics[height=6cm]{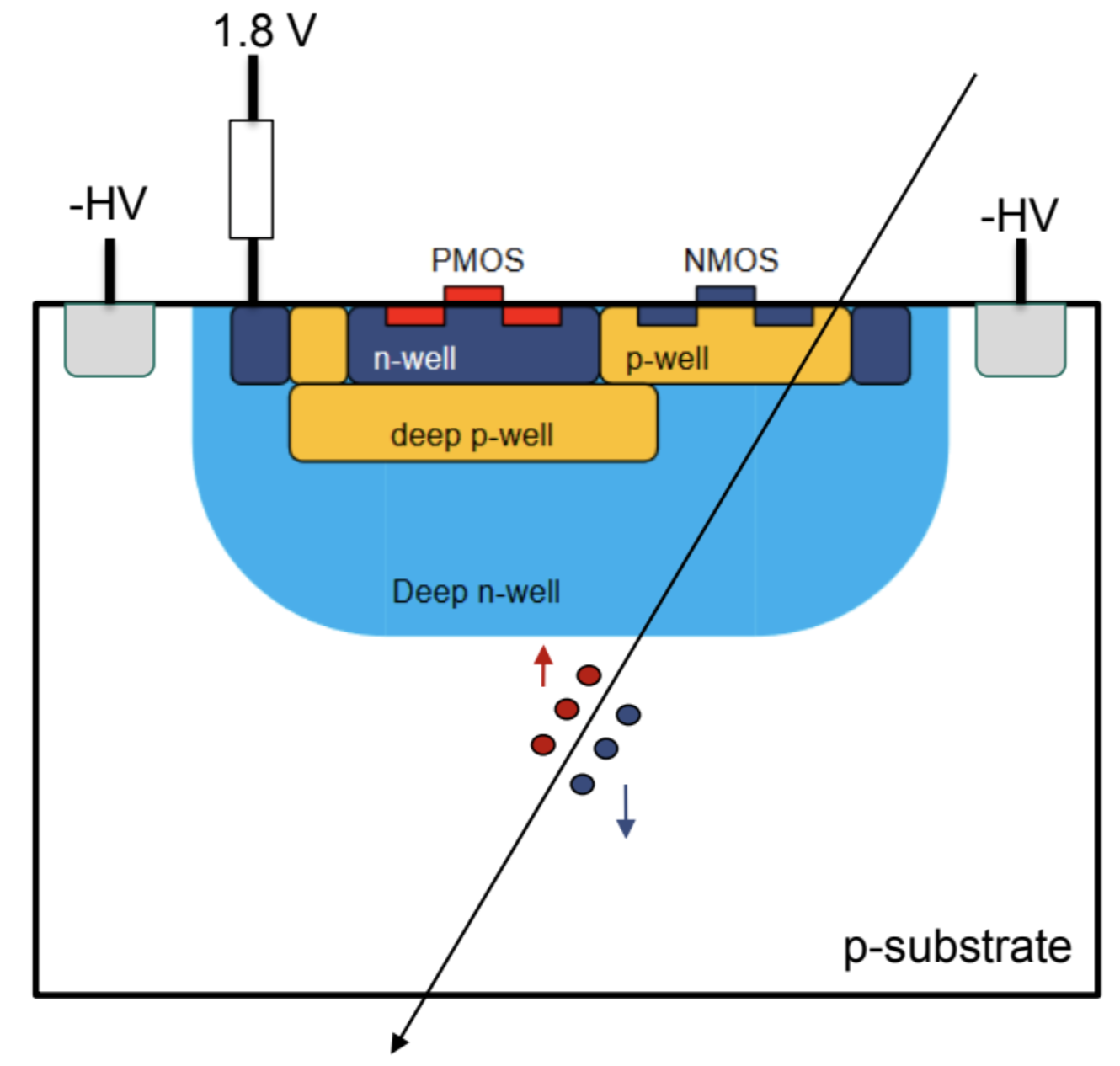}&& b) \includegraphics[height=4cm]{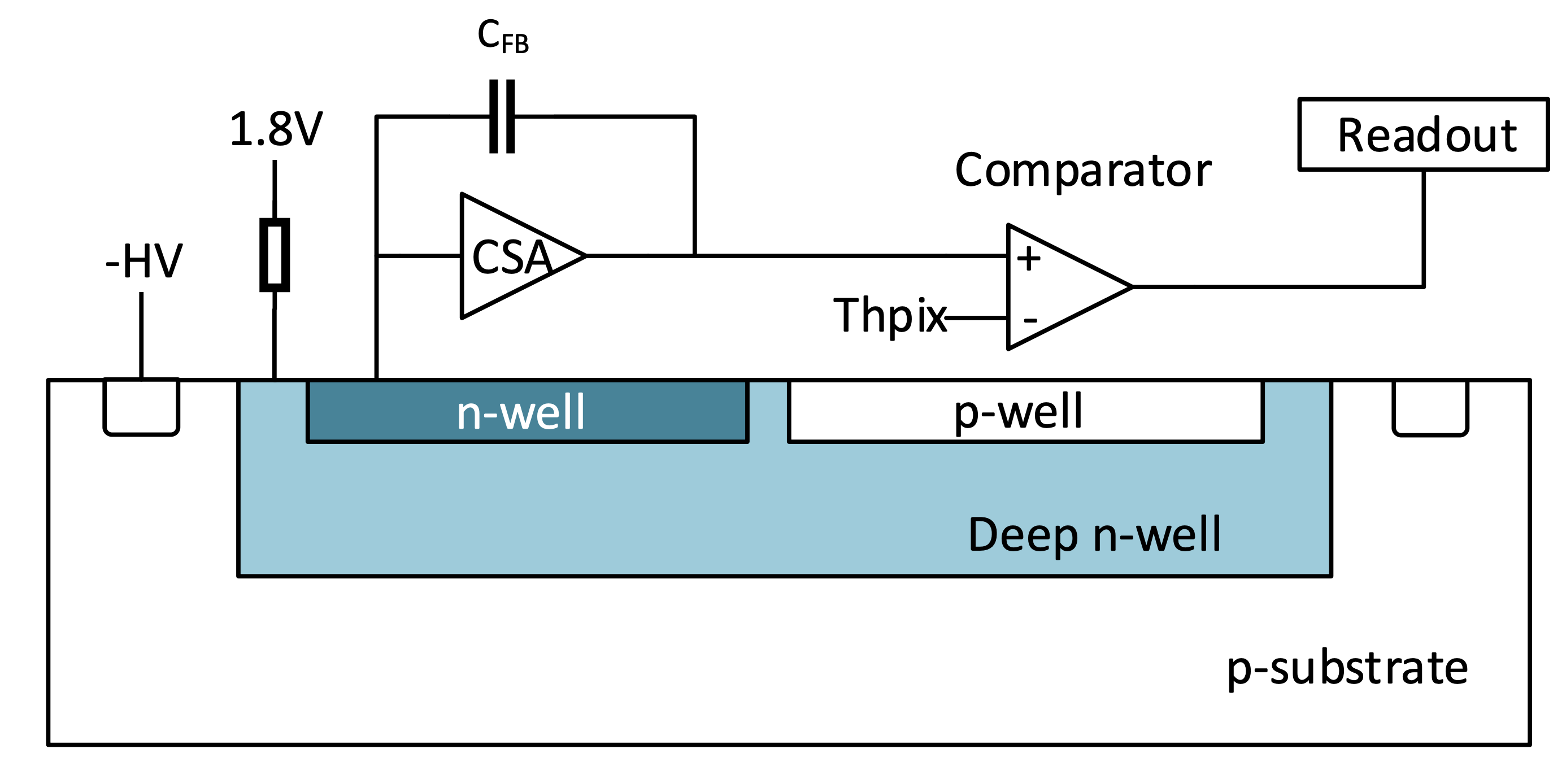}
   \end{tabular}
   \end{center}
   \caption{ \label{fig:cmos} 
        AstroPix is implanted in HVCMOS technology with a $500~\times~500~\mu\mathrm{m}^2$ pixel pitch. a) CMOS circuitry (gold and dark blue) is implanted directly into the surface of the AstroPix substrate inside a shielding 6~nm n-well (light blue). The high-voltage contacts (gray) are implemented at the pixel edges. b) The pixel circuit enables signal amplification and self triggering via a comparator before sending the data to a digital periphery on-chip. This eliminates the need for an external readout ASIC.}
\end{figure}

AstroPix is a pixelated silicon HVCMOS sensor designed for future application in space-based gamma-ray astronomy and nuclear physics. The CMOS processing (see Fig.~\ref{fig:cmos}) enables charge collection, signal amplification, and self-triggering independently in every pixel. Charge collection efficiency is enhanced with the application of a high-voltage bias\cite{Peric:2018lya} delivered to every pixel. The resulting signal from each pixel is ultimately read out to a digital periphery on the bottom of the chip which fully digitizes hits from the entire array before returning an encoded bitstream without the need for an external ASIC. A deeper discussion of the functionality of AstroPix can be found in Refs.~\citenum{Brewer:2021mbe} and \citenum{spie}.

\begin{figure} [ht]
   \begin{center}
   \begin{tabular}{c} %% tabular useful for creating an array of images 
   \includegraphics[height=6cm]{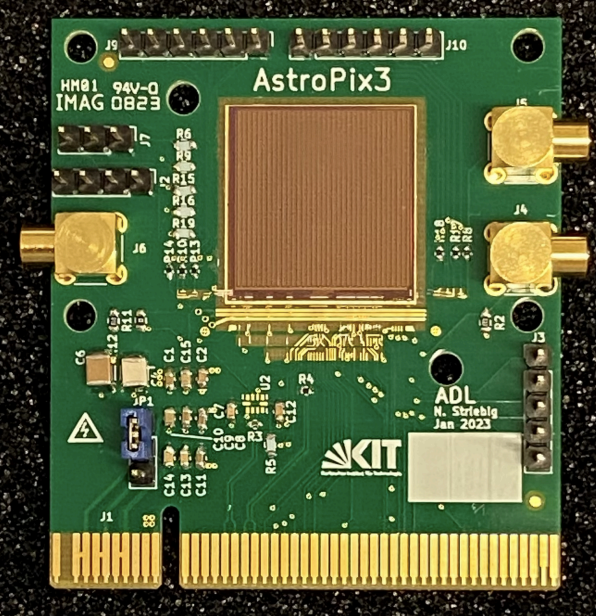}
   \end{tabular}
   \end{center}
   \caption{ \label{fig:v3Board} 
        The $2\times2$~cm$^2$ AstroPix\_v3 chip is mounted on a custom-designed Printed Circuit Board (PCB) for bench testing. High voltage is delivered on the left side and analog signals off select pixel amplifiers are read off from the bottom right.}
\end{figure}

The chip design development involves multiple iterations of chips with incremental changes in order to step the technology closer to its ultimate design requirements. The current version under test is AstroPix version 3 or AstroPix\_v3, featuring the final pixel size of $500\times500~\mu\mathrm{m}^2$ (see Fig.~\ref{fig:v3Board}).

AstroPix requirements are inherited from the All-sky Medium Energy Gamma ray Observatory eXplorer, AMEGO-X\cite{amx_paper}, a medium energy gamma ray (25 keV - 1 GeV) mission concept. The science case of this larger instrument is enabled by the use of AstroPix in its tracking system rather than heritage silicon strip technology. The final AstroPix chip is also slated to be utilized in the Barrel Imaging Calorimeter of the ePIC dectector at the future Electron-Ion Collider (EIC). 

AstroPix pixels measure $500\times500~\mu\mathrm{m}^2$ in pitch. To minimize the risk of parasitic couplings, the n-well implant in each pixel which protects the CMOS electronics is decreased to $300\times300~\mu\mathrm{m}^2$ within the $500\times500~\mu\mathrm{m}^2$ pixel pitch. This allows for increased space between the n-well and the high-voltage contacts.

To achieve the science desired by AMEGO-X and the EIC, AstroPix must achieve a dynamic range of 25 - 700 keV per pixel. The high end of this range is enabled with the thick 500~$\mu$m wafer, which must be fully depleted. In other words, the applied high-voltage bias must induce an active charge collection area into the silicon bulk to a depth of 500~$\mu$m. This depletion depth, in cm, is defined as
\begin{equation}
    \label{eq:depl}
    d = \sqrt{\frac{2\epsilon_{\mathrm{Si}}}{q}\frac{V_{\mathrm{bias}}}{N_{\mathrm{eff}}}}~,
\end{equation}
where $\epsilon_{\mathrm{Si}} = 1.03 \times10^{-12}$~C$*$cm/V is the permitivitiy of silicon, $q = 1.60\times10^{-19}$~C is the elementary charge, $V_{\mathrm{bias}}$ is the applied high-voltage bias, and $N_{\mathrm{eff}}$ is the silicon p-doping concentration. The doping concentration is related to the silicon wafer resistivity like 
\begin{equation}
    \label{eq:neff}
    N_{\mathrm{eff}} = \mathrm{exp}\left[43.28 - \mathrm{log}(\rho) - \mathrm{log}\left(\frac{482.8}{1 + 0.1322\rho^{-0.811}}\right)\right]~,
\end{equation}
where $\rho$ is the wafer resistivity\cite{depl_calc} in units of $\Omega*$cm. Therefore, an estimation of depletion depth can be made with knowledge of the wafer resistivity and the applied high-voltage bias.

\begin{figure} [ht]
   \begin{center}
   \begin{tabular}{c} %% tabular useful for creating an array of images 
   \includegraphics[height=6cm]{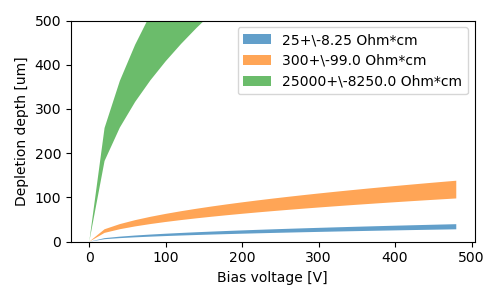}
   \end{tabular}
   \end{center}
   \caption{ \label{fig:depl_theory} 
        The depletion of a silicon wafer, as described in Eq.~(\ref{eq:depl}), is shown for a variety of silicon resistivities as a function of applied high-voltage bias. This illustrates the theoretical depth of depletion that AstroPix\_v3 might expect to achieve.}
\end{figure}

Achieving 500~$\mu$m depletion is a balance of applied high-voltage bias and bulk silicon resistivity. The space environment limits the maximum applied voltage, and cost and fabrication constraints can limit resistivity. Therefore, AstroPix plans to achieve depletion with a -400~V high voltage and bulk resistivity around 5~k$\Omega*$cm. Figure~\ref{fig:depl_theory} shows the growth of depletion depth as a function of applied bias voltage as given by Eq.~\ref{eq:depl}. Lower resistivity wafers with $25\pm8.25$ and $300\pm100~\Omega*$cm cannot achieve the desired depletion depth within the given bias range, but these lower-resistivity wafers are more straightforward to obtain, fabricate, and test.

%%%%%%%%%%%%%%%%%%%%%%%%%%%%%%%%%%%%%%%%%%%%%%%%%%%%%%%%%%%%%%%%%%%%%%%%%%%%%%%%%%%%%%%%%%%%%%%%%%%%%%%%%%%%%%%%%%%%%%%%%%%%%%%%%%%%%%%%%%%%%%%%%%%%%%%%%%%%%%%%%%%%%%%%%%%%%%%%%%%%%%%%%%%%%%%%%%%%%%%%%%%%%%%%%%%%%%%%%%%%%%%%%%%%%%%%%%%%%%%%%%%%%%%%%%%%%%%%%%%%%%%%%%%%%%%%%
\section{HIGH-RESISTIVITY WAFER TESTING}
\label{sec:highres}

Chips manufactured on three wafer resistivities of AstroPix\_v3 were tested for their electrical properties and data quality. The relationship between applied bias voltage and the subsequent leakage current off the high-voltage line is of particular importance. This correlation is illustrated on current-voltage or IV curves (methodology is explored in Section~\ref{ssec:hr_method}). The expected nature of these curves, as a function of bias voltage, involves a long sloping section of low leakage until an exponential rise. The inflection point indicates the breakdown voltage of the chip, or the maximum voltage that can be applied before potential harm to the electronics and data quality. AstroPix\_v3 is designed for a breakdown voltage of -400V, and expected to draw $<1~\mu$A of leakage current over its operational range. This value enables operation in systems with minimal cooling structures and efficient power delivery. 

\begin{figure} [ht]
   \begin{center}
   \begin{tabular}{ccc} %% tabular useful for creating an array of images 
   a) \includegraphics[height=6cm]{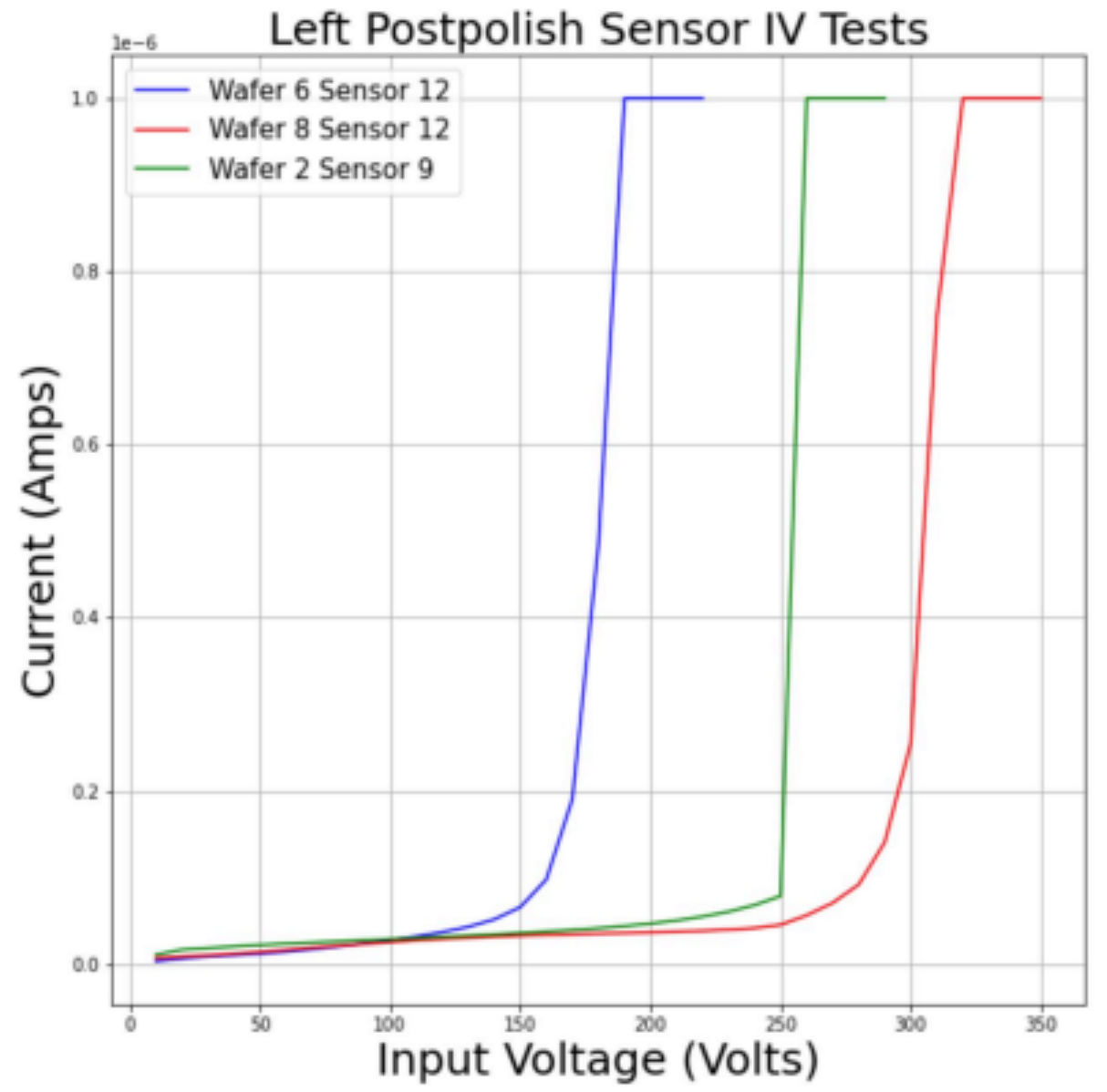}&& b) \includegraphics[height=6cm]{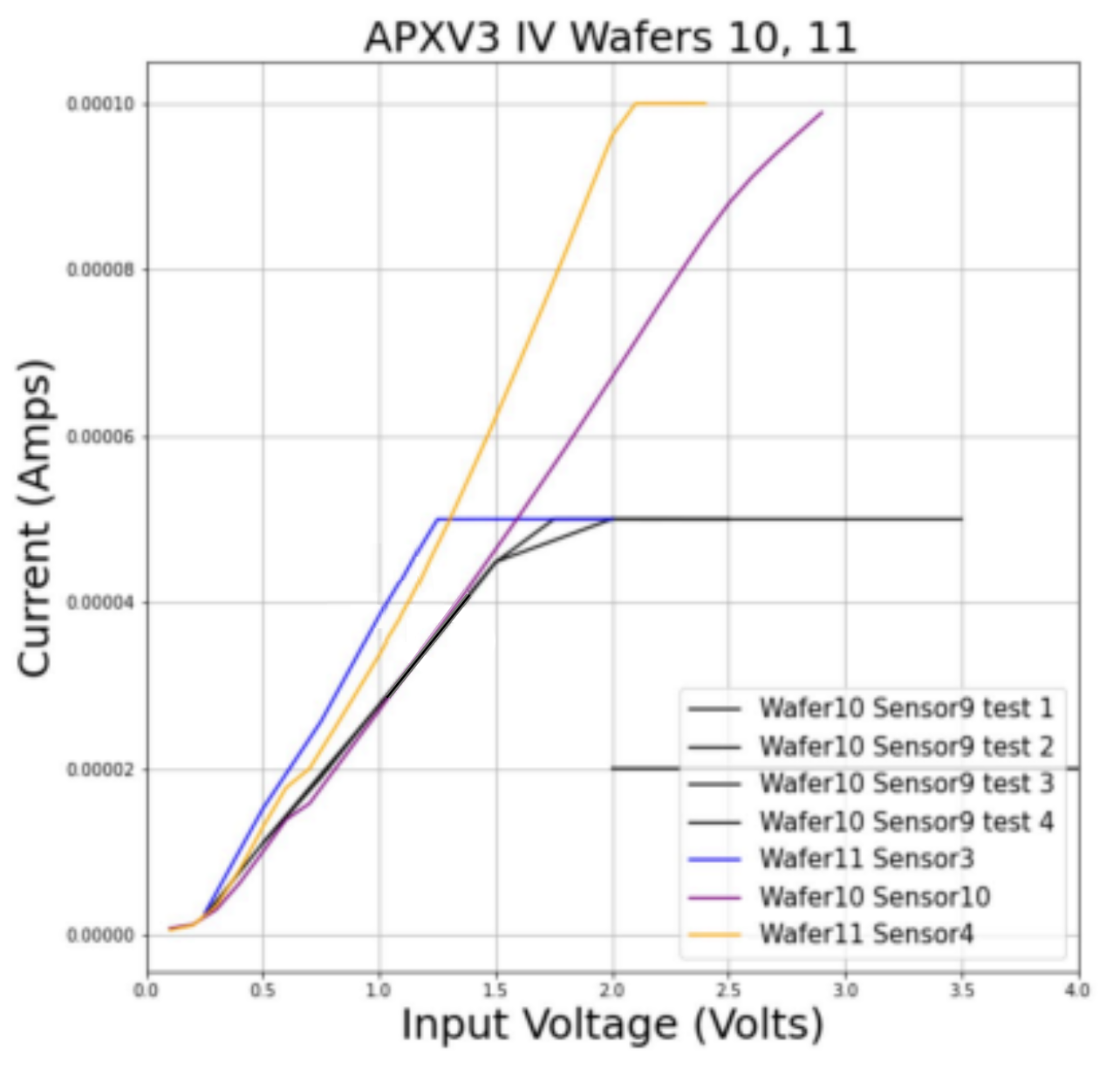}
   \end{tabular}
   \end{center}
   \caption{ \label{fig:iv_orig} 
        The current-voltage (IV) relationship of AstroPix chips fabricated on different resistivity wafers shows breakdown and high leakage currents at different values. The chips are designed for breakdown at -400V. a) Low- and medium- resistivity AstroPix\_v3 chips ($25\pm8.25~\Omega*$cm of wafer 2 and $300\pm100~\Omega*$cm of wafers 6 and 8) show expected behavior with exponential leakage current draw beginning in the hundreds of volts. b) The high-resistivity wafers 10 and 11 ($25000\pm 8250~\Omega*$cm) exhibit linear behavior. The leakage current exceeds 40~$\mu$A with -2~V applied which is much higher than the benchmark $1~\mu$A leakage at full depletion.}
\end{figure}

The low- and medium-resistivity wafers ($25\pm8.25~\mathrm{and}~300\pm100~\Omega*$cm, respectively) reflect the expected IV curve shape (see Fig.~\ref{fig:iv_orig} a)) with measured breakdown voltages between -150 - -290V. The high-resistivity wafer ($25000\pm8250~\Omega*$cm) (see Fig.~\ref{fig:iv_orig} b)) creates an IV curve that is linear and therefore unexpected. The scale is also much smaller, seeing this linear leakage draw with only a few volts of applied bias. The benchmark $1~\mu$A leakage current is surpassed with $<0.2$~V bias. In this case, no breakdown voltage can be defined.

Operation and data quality of the high-resistivity chips also suffered. Pixels around the edge of the array were notably noisier than pixels at the center, indicating the possibility of some edge effects. All pixels on these high-resistivity chips were noisier than the average pixel on a low- or medium-resistivity wafer, in part due to the inability to bias effectively. 
 
%%%%%%%%%%%%%%%%%%%%%%%%%%%%%%%%%%%%%%%%%%%%%%%%%%%%%%%%%%%%%%%%%%%%%%%%%%%%%%%%%%%%%%%%%%%
\subsection{Methods}
\label{ssec:hr_method}

IV curves for AstroPix chips have been measured at the Santa Cruz Institute for Particle Physics (SCIPP) at the University of California Santa Cruz and at Argonne National Laboratory with consistent results. IV measurements can be performed on wafers before the dicing of individual chips, on bare chips, and on chips mounted to custom-designed Printed Circuit Boards (PCBs). Wafer-level testing and bare chip testing is conducted within a probe station, where probe tips target high-voltage delivery and return pads present on each chip. Full power is not delivered to the chip, and data collection is not possible. The probe tips can measure the delivered voltage and subsequent current directly from these pads. Testing of mounted chips (on PCBs) involves the delivery of high voltage to the board which is then routed to the chip via wire bonds to HV pads. The supplied voltage and leakage current is the measured from the power supply or source meter providing the input power. Mounted chips can be fully operated, with all power lines applied and data collection before, during, and after the test is possible. 

A thermal investigation of the chips was conducted at SCIPP with supporting effort from NASA Goddard Space Flight Center (GSFC). Both topside and backside of the chip were imaged with a thermal camera utilizing wavelengths of $7.5-13~\mu$m. The bulk silicon is transparent at these wavelengths. A magnifying lens is used to obtain higher resolution images of structures on the pixel-level.

%%%%%%%%%%%%%%%%%%%%%%%%%%%%%%%%%%%%%%%%%%%%%%%%%%%%%%%%%%%%%%%%%%%%%%%%%%%%%%%%%%%%%%%%%%%
\subsection{Results}
\label{ssec:hr_results}

\begin{figure} [ht]
   \begin{center}
   \begin{tabular}{c} %% tabular useful for creating an array of images 
   \includegraphics[height=5cm]{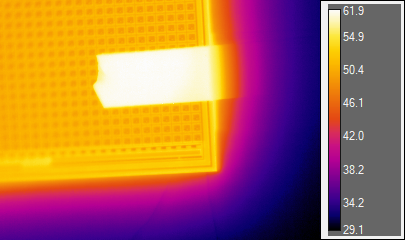}
   \end{tabular}
   \end{center}
   \caption{ \label{fig:thermal} 
        The face of an AstroPix\_v3 high-resistivity chip as imaged with a thermal camera. The horizontal wide band is an artifact from mounting tape. When high voltage is applied, the chip edge (in the bottom-right of the image) gets hotter than the rest of the chip. The thermal gradient across the chip indicates that a higher power density is generated on the chip edge unexpectedly.}
\end{figure}

Backside illumination of high-resistivity AstroPix\_v3 chips revealed temperature gradients across the chip as a whole and within individual pixels. Without bias voltage applied, individual pixels and their central implants can be seen where the central implant region appeared hotter than the guard structure around the implant by approximately 1$^o$~C. This is likely due to emissivitiy differences. When the bias voltage was applied, the gradient inverted and the central implants register cooler than the surrounding area by more than 1$^o$~C. The chip periphery region (around the edge) is also more than $4^o$~C warmer than the pixel matrix on average (see Fig.~\ref{fig:thermal}). The temperature across the pixel array is consistent (intra-pixel variation is consistent but no trends between pixels is noted). In general, the temperature gradient is ordered like $T_{\mathrm{chip~edge}}>T_{\mathrm{periphery}}>T_{\mathrm{pixel~edge}}>T_{\mathrm{pixel~implant}}$. Topside and backside measurements yielded similar results.

IV measurements of high-resistivity AstroPix\_v3 chips were inconsistent depending on the method of measurement. The linear curve with very high currents shown in Figure~\ref{fig:iv_orig} b) was measured with a chip in a probe station, but measurements of the same chip mounted on the PCB showed even higher leakage currents. This lead to reconsideration of the PCB design and mounting strategy. 

\begin{figure} [ht]
   \begin{center}
   \begin{tabular}{ccc} %% tabular useful for creating an array of images 
   \includegraphics[height=7cm]{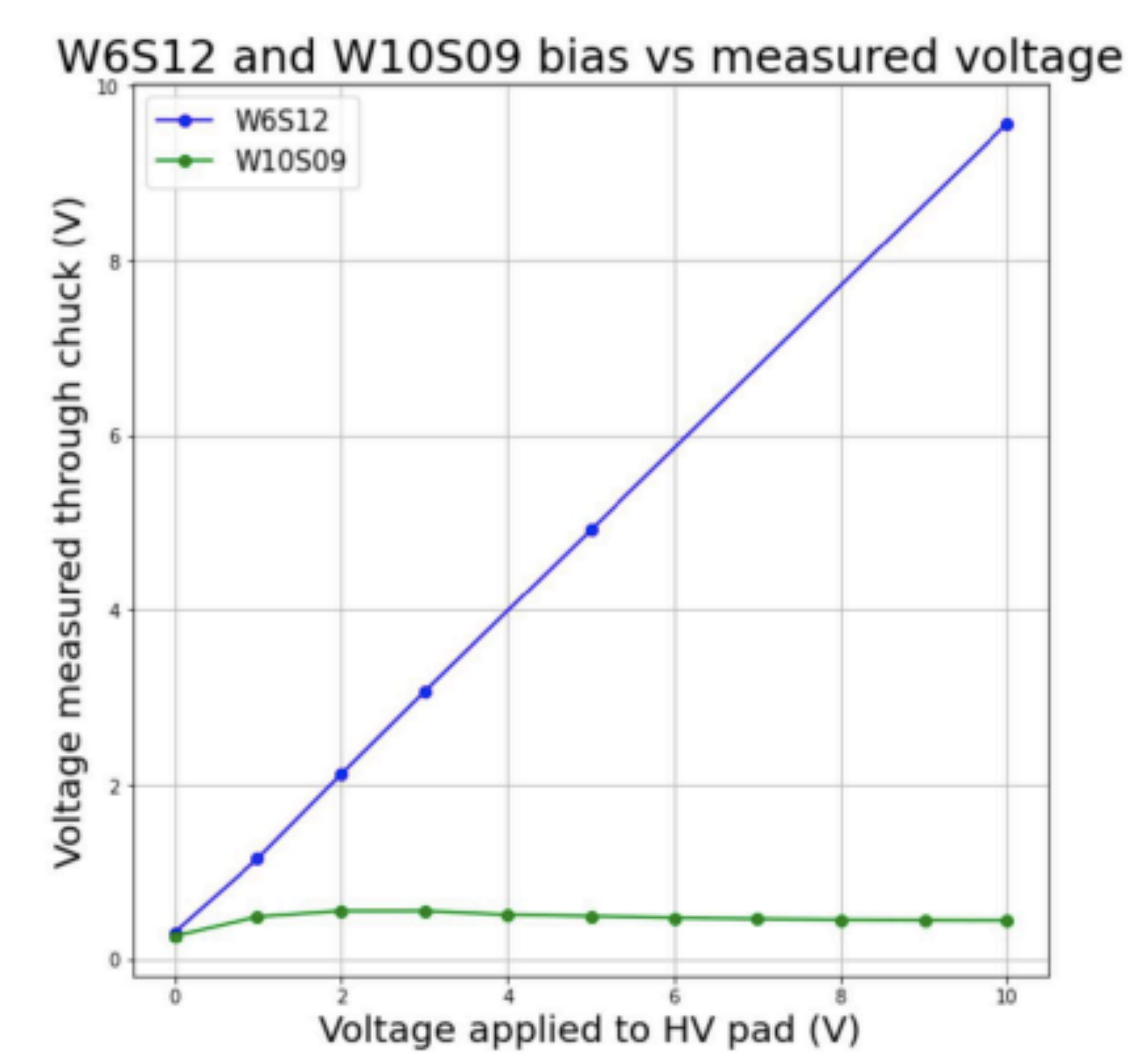} 
   \end{tabular}
   \end{center}
   \caption{ \label{fig:v_back} 
        The relationship between delivered bias voltage and measured backside voltage differ between the medium-resistivity (wafer 6) and high-resistivity (wafer 10) AstroPix\_v3 chips. The medium-resistivity displayed expected, linear behavior however the high-resistivity chip which draws high leakage current shows only a very small backside potential.}
\end{figure}

Images of the HV pads housing wirebonds from the chip to the PCB indicate no damage or shorts, indicating that the chip is not damaged by bonding. Backside voltage was measured on bare medium- and high-resistivity chips (see Fig.~\ref{fig:v_back}). The medium-resistivity (wafer 6) chip shows an expected linear relationship between applied voltage and backside measured voltage, however the high-resistivity chip shows only a very small potential indicating different behavior on the backside of the high-resistivity chips.

This voltage gradient on the backside of the chips is likely an inherent silicon wafer property. Mounting of bare chips on the PCB uses conductive tape in order to ensure a good electrical ground. However, the inherent backside potential of AstroPix was likely enhanced with the conductive tape leading to the observed high leakage currents. To test this hypothesis, double-sided Kapton tape was applied under the chip for insulation from the PCB.

\begin{figure} [ht]
   \begin{center}
   \begin{tabular}{c} %% tabular useful for creating an array of images 
   \includegraphics[height=7cm]{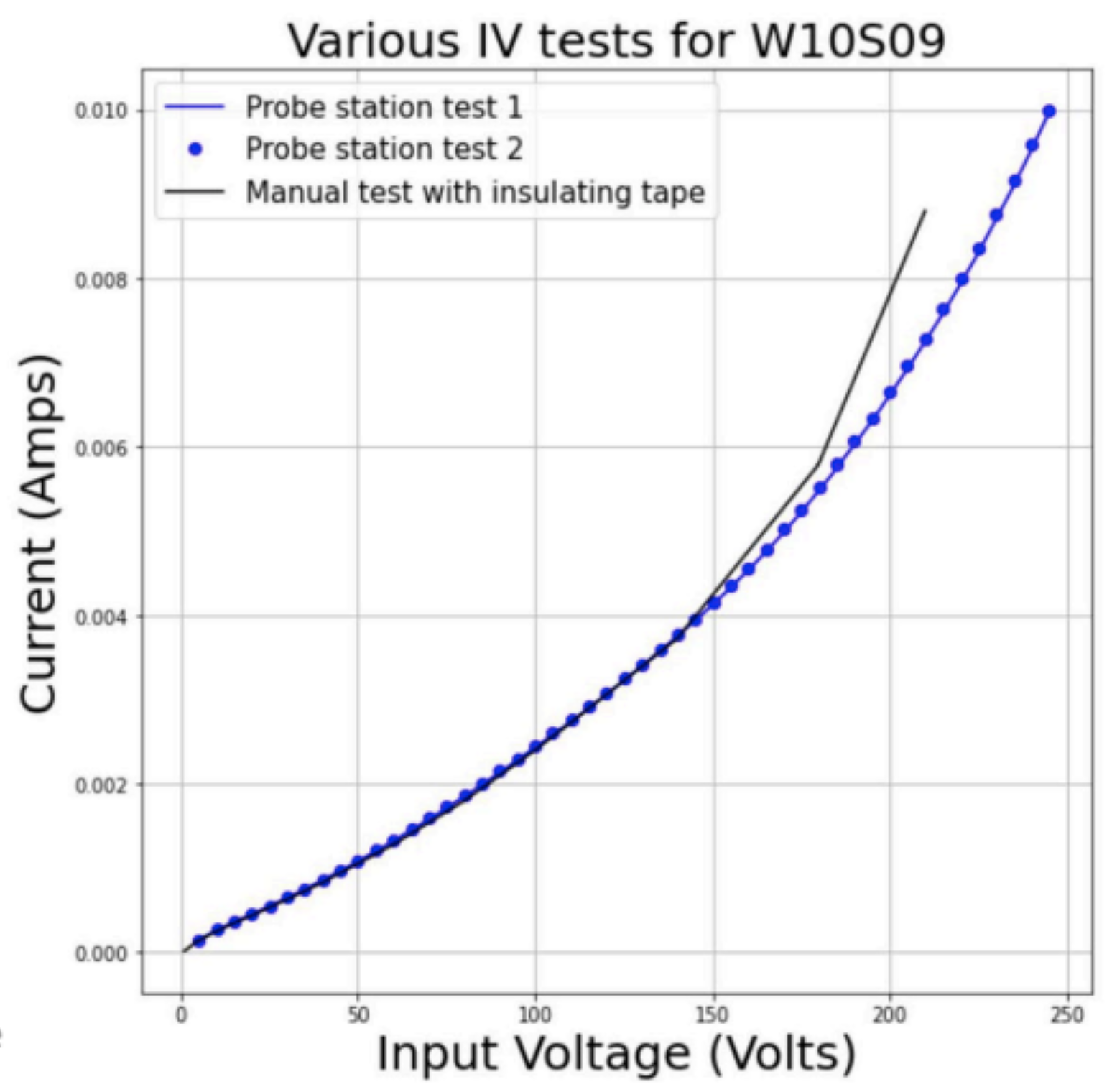}
   \end{tabular}
   \end{center}
   \caption{ \label{fig:iv} 
        The relationship of the leakage current off the biasing high-voltage line and the delivered voltage (blue data points) show robust behavior when probed as a bare silicon die before mounting to the PCB. b). After mounting the same chip to a PCB with insulating tape rather that conductive tape, the current/voltage curve mimics the probe station measurement, indicating that the conductive tape previously used to mount the chips to the PCB caused the problematic high leakage.}
\end{figure}

The resulting IV curve with insulating mounting (see Fig.~\ref{fig:iv}) does replicate the behavior of the bare chip before mounting. This indicates that the high leakage current previously measured was due in some part to an inherent voltage gradient on the backside of the chip which was shorted with the use of conductive mounting. Insulating the silicon backside from the PCB then enables the delivery of larger bias voltages.

After this remounting, thermal images were collected again. The same temperature features as shown in Fig.~\ref{fig:thermal} were still present

AstroPix required $500~\mu$m depletion can only be achieved with a relatively high-resistivity ($>5\mathrm{k}~\Omega*$cm) wafer, so the behavior of the AstroPix\_v3 high-resistivity wafer must be understood to inform future versions. Some hypotheses regarding the measured behavior include:
\begin{itemize}
    \item{The very high currents of the high-resistivity chip mounted on the carrier board was due to the inherent voltage gradient on the backside and the conductive tape used in the mounting (as explored above).}
    \item{Long minority lifetimes in the silicon (8~ms) could contribute to the high leakage currents as the diffusion of these extra electrons/holes generated on the chip sides or backplane (5~mm diffusion length) would increase the measured current.}
    \item{Power is generated on the chips edge/periphery. This could impact the poorer data quality of edge pixels and induce edge surface currents which impact/saturate pixel amplifier performance.}
    \item{A self-quenching effect may occur if electrons are always leaving the un-depleted bulk. This may lead to the net positive charge cloud that captures their drift.}
\end{itemize}

%%%%%%%%%%%%%%%%%%%%%%%%%%%%%%%%%%%%%%%%%%%%%%%%%%%%%%%%%%%%%%%%%%%%%%%%%%%%%%%%%%%%%%%%%%%
\subsection{Next Steps}
\label{ssec:hr_steps}

The improved IV behavior of the high-resistivity chips mounted with insulating tape is encouraging and warrants further data quality testing. Special attention will be paid to the noise nature of the pixels to gauge whether the noisier edge-effect is still present. The long-lifetime hypothesis will be studied with detailed tests of the biasing structures and repeating these studies on other high-resistivity sensors.

While a step in the right direction, the improved IV curve still does not satisfy AstroPix design benchmarks of $<1~\mu$m leakage with a -400~V bias. Continued design and fabrication tests will be performed with future versions of AstroPix on similarly high-resistivity silicon wafers to monitor the performance as the chip design progresses (see Section~\ref{sec:concl}).

%%%%%%%%%%%%%%%%%%%%%%%%%%%%%%%%%%%%%%%%%%%%%%%%%%%%%%%%%%%%%%%%%%%%%%%%%%%%%%%%%%%%%%%%%%%%%%%%%%%%%%%%%%%%%%%%%%%%%%%%%%%%%%%%%%%%%%%%%%%%%%%%%%%%%%%%%%%%%%%%%%%%%%%%%%%%%%%%%%%%%%%%%%%%%%%%%%%%%%%%%%%%%%%%%%%%%%%%%%%%%%%%%%%%%%%%%%%%%%%%%%%%%%%%%%%%%%%%%%%%%%%%%%%%%%%%%
\section{EDGE TRANSITION CURRENT TECHNIQUE TEST FOR DIRECT DEPLETION MEASUREMENT}
\label{sec:etct}

The full AstroPix dynamic range is only accessible when full depletion of 500~$\mu$m is achieved. Initial studies with gamma-ray sources use an indirect means of estimating the depletion depth of the sensor\cite{spie_suda} by considering the measured count rate\cite{pixel}. These studies found that the shape and amplitude of depletion as a function of applied bias voltage agreed with theory when considering the large uncertainty on resistivity. The indirect measurement nature reflects charge collection efficiency and potential charge sharing which could be secondary effects to the depletion depth itself. Therefore, a direct measurement was deemed necessary.

%%%%%%%%%%%%%%%%%%%%%%%%%%%%%%%%%%%%%%%%%%%%%%%%%%%%%%%%%%%%%%%%%%%%%%%%%%%%%%%%%%%%%%%%%%%
\subsection{Method}
\label{ssec:etct_method}

To more directly probe AstroPix depletion, an edge Transition Current Technique (eTCT)\cite{etct} was used at SCIPP. This strategy utilizes an infrared laser with 15~$\mu$m spot size to illuminate the edge of a diced AstroPix chip. The laser is scanned through the thickness of the chip from the implanted surface down to the mounted backside, and analog data (directly off the amplifier) off the edge pixel is collected. It is expected that when the laser scans outside of the active depletion area, then the analog response will lessen due to the recombination process allowing for a direct measurement of the spatial extent of the depletion region. 

AstroPix\_v3 chips of all resistivities ($25\pm8.25,~300\pm100,~\mathrm{and} 25000\pm8350~\Omega*$cm,) were tested. The eTCT laser setup features a remote-controlled stage with 1~$\mu$m resolution and a 1016~nm wavelength infrared laser. The laser is pulsed at 1~kHz (or 5~kHz for the high-resistivity chip) and a trigger from this system is used to collect AstroPix response data with each laser pulse even if the integrated comparator did not record digital data. The laser intensity is tunable and was varied based upon the needs of each different AstroPix resistivity (from 2-8\%). 

AstroPix\_v3 chips were all mounted on standard testing PCBs (as in Fig.~\ref{fig:v3Board}) without populated connectors as to avoid potential reflections in the system. However, the chip edge is still 2~cm away from the edge of the board. To avoid reflections, one diced edge of the AstroPix chips was polished. However due to the inverted mounting of the chip PCB to the stage, this polished edge did not face the laser. In other words, the laser was incident upon a non-polished edge of the diced AstroPix array. The chip and laser were aligned by eye and with the help of a 640~nm red laser. The center of the incident pixel was identified from preliminary scans across large areas of silicon in all three dimensions in order to identify laser waist depth into the silicon and orient the laser within one single pixel. This alignment strategy introduces large systematic errors in alignment including potential relative angles between the laser propagation direction and the AstroPix edge. 

\begin{figure}[ht]
   \begin{center}
   \begin{tabular}{c} %% tabular useful for creating an array of images 
   \includegraphics[height=10cm]{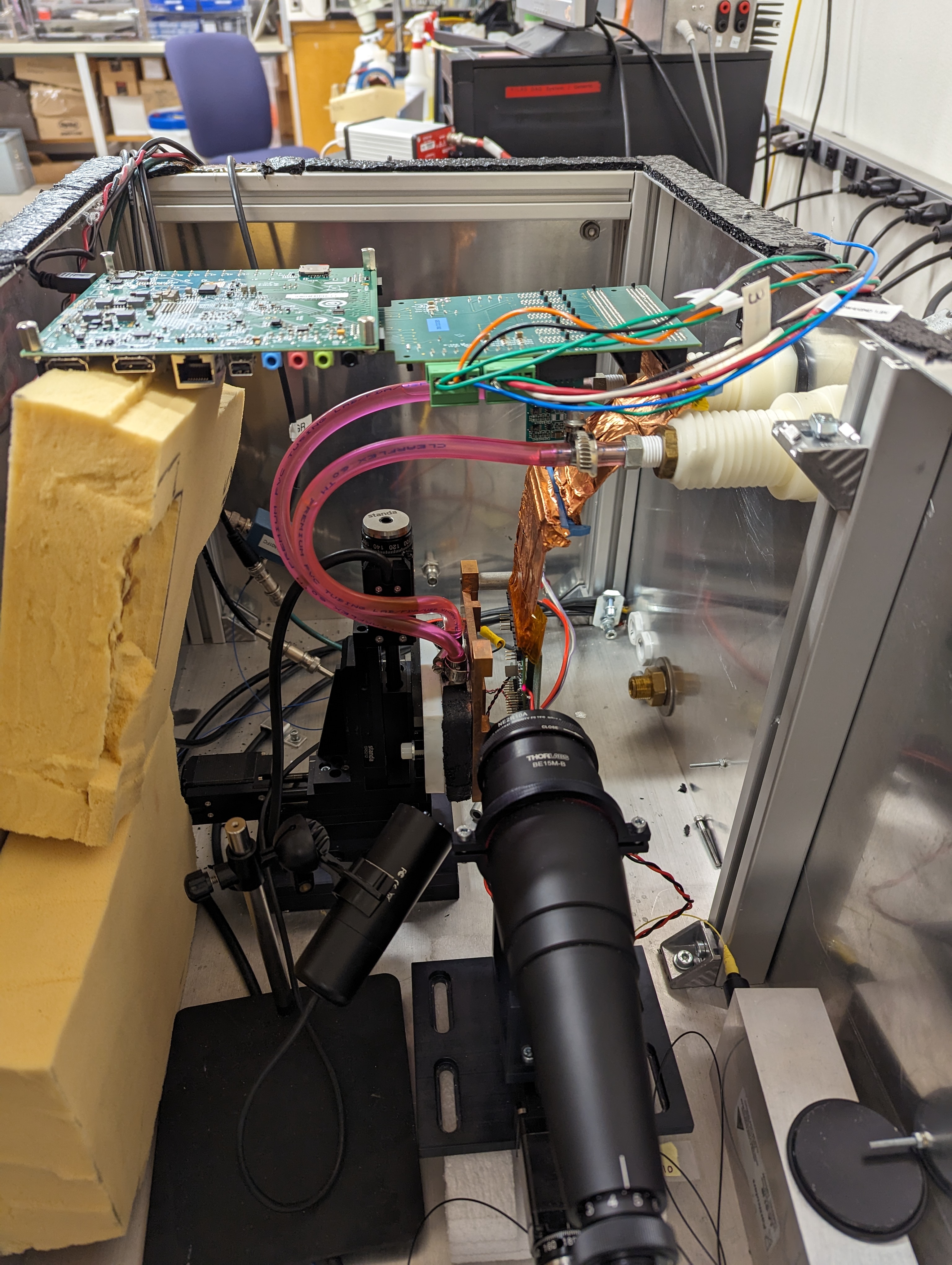}
   \end{tabular}
   \end{center}
   \caption{ \label{fig:etct} 
        The edge-TCT technique requires scanning the edge of the AstroPix chip. From the black laser lens at the bottom of the figure, a red beam spot can be seen on the inverted AstroPix chip indicating the location of the laser. The AstroPix chip is connected to a remote-controlled stage with 1~$\mu$m resolution, allowing the chip bulk to move through the focused laser. Scanning the response of AstroPix to the same laser signal at different depth of the pixel provides a direct measurement of depletion.}
\end{figure}

The laser lens itself remained stationary for all testing, and the AstroPix chip under test was mounted to the movable stage (see Fig.~\ref{fig:etct}). Supporting electronics including an FPGA are inverted to allow the chip on a 4~in ribbon cable to hang in the path of the laser. In this upside-down orientation, pixels in column 0 are in the direct path of the laser, however only row 0 supports analog data readout so the corner pixel in (row 0, column 0) was targeted for testing. Analog data was collected on an oscilloscope that collected 100 individual waveforms for each laser scan location. Bulk analysis of this data (in Sec.~\ref{ssec:etct_results}) considers one averaged waveform from each scan point. A taller analog pulse height correlates to a larger measured energy. 

\begin{figure} [ht]
   \begin{center}
   \begin{tabular}{c} %% tabular useful for creating an array of images 
   \includegraphics[height=3cm]{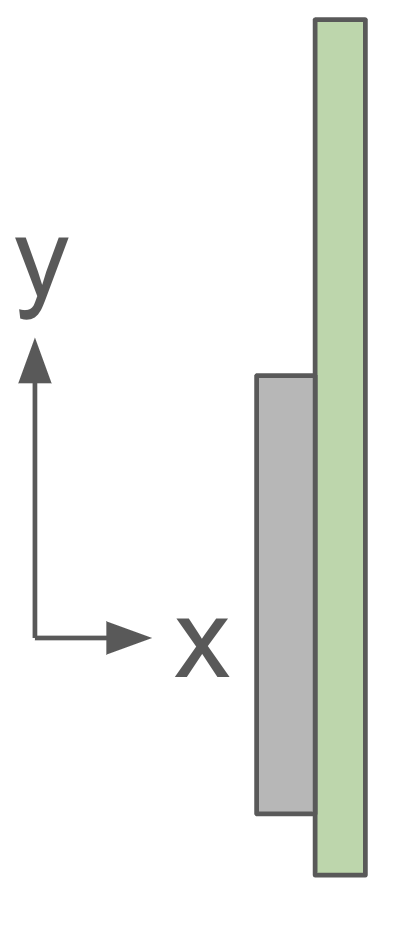}
   \end{tabular}
   \end{center}
   \caption{ \label{fig:coords} 
        eTCT system coordinates are defined by motion of the stage and AstroPix chip. The $z$ dimension is the same as the laser propagation direction, and increasing $z$ moves the laser spot further into the bulk of the silicon. The $xy$ plane defines the edge of the diced AstroPix chip. To begin a scan, the laser location was placed at a relatively low $x$ value to the left of the chip surface. As the stage $x$ dimension was increased, the laser interacted first with the chip surface and ultimately with the chip backside mounted to the PCB. The $y$ dimension moves the chip up and down to move the laser to the extremes of the pixel pitch, with positive $y$ moving the laser toward the chip periphery.}
\end{figure}

One chip of each resistivity was tested in this identical setup. Laser settings including the frequency and intensity were tuned for best performance (highest pixel response) for each substrate. Identical scans were conducted for a range of high-voltage biases, also as appropriate for each resistivity. Absolute stage locations were duplicated between runs, however potential systematic offsets  by the tension of wires, incomplete stage motion, or the settling of the system may have been introduced and were not tracked. The coordinate system (see Fig.~\ref{fig:coords}) defines $z$ as the direction of laser propagation and the $xy$ plane as the AstroPix chip edge. The laser is held stationary for every test and the chip mounted to the movable stage is moved, meaning that increasing $x$ scans through the laser beam from the surface of the chip through the bulk to the chip backside that is mounted on the PCB. The $y$ dimension is used to scan the height of the pixel pitch with positive $y$ moving closer to the chip periphery at the bottom of the array.  

%%%%%%%%%%%%%%%%%%%%%%%%%%%%%%%%%%%%%%%%%%%%%%%%%%%%%%%%%%%%%%%%%%%%%%%%%%%%%%%%%%%%%%%%%%%
\subsection{Results}
\label{ssec:etct_results}

Limited conclusions can be drawn from the testing of a high-resistivity chip, as this eTCT testing was preformed before the chip mounting strategy was updated to utilize insulating tape. The high leakage current and poor data quality of edge pixels (as explored in Sec.~\ref{sec:highres}) only allowed a bias scan from -(2-10)~V. Individual waveforms from these tests do not behave as expected, where pulses in response to higher bias voltage have lower pulse duration and amplitude than those with lower bias voltages. This could be due to the FE amplifier saturation with high leakage current present even at these low voltages. No conclusions can therefore be drawn regarding the depletion of chips with $25000\pm8250~\Omega*$cm (high resistivity). 

Additionally, few conclusions can be drawn from the medium-resistivity $300\pm100~\Omega*$cm test chip, measured with bias voltages ranging from -(50-400)~V, due to experimental shortcomings. The chip originally intended for testing did not respond reliably to standard data acquisition software, so an alternative chip with the same resistivity was used. However, the alternate chip was mounted on a standard PCB (see Fig.~\ref{fig:v3Board}) with all connectors still mounted to the board. The presence of these additional metallic surfaces very close to the lasers propagation path toward the bottom left pixel of (row 0, column 0) led to the measurement of very large reflections. These reflections complicate the analysis of these data. This analysis is still underway but it is expected that that the complicating reflections will yield large measurement uncertainties.

\begin{figure} [ht]
   \begin{center}
   \begin{tabular}{ccc} %% tabular useful for creating an array of images 
   a) \includegraphics[height=5cm]{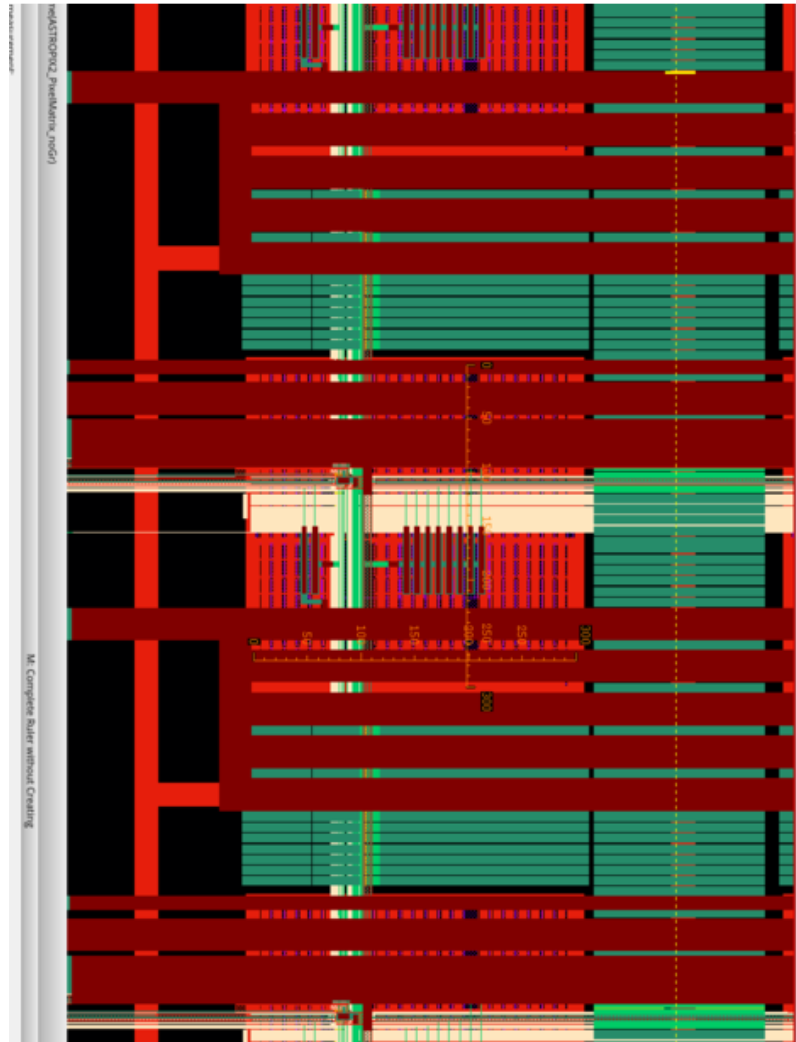}&&b) \includegraphics[height=5cm]{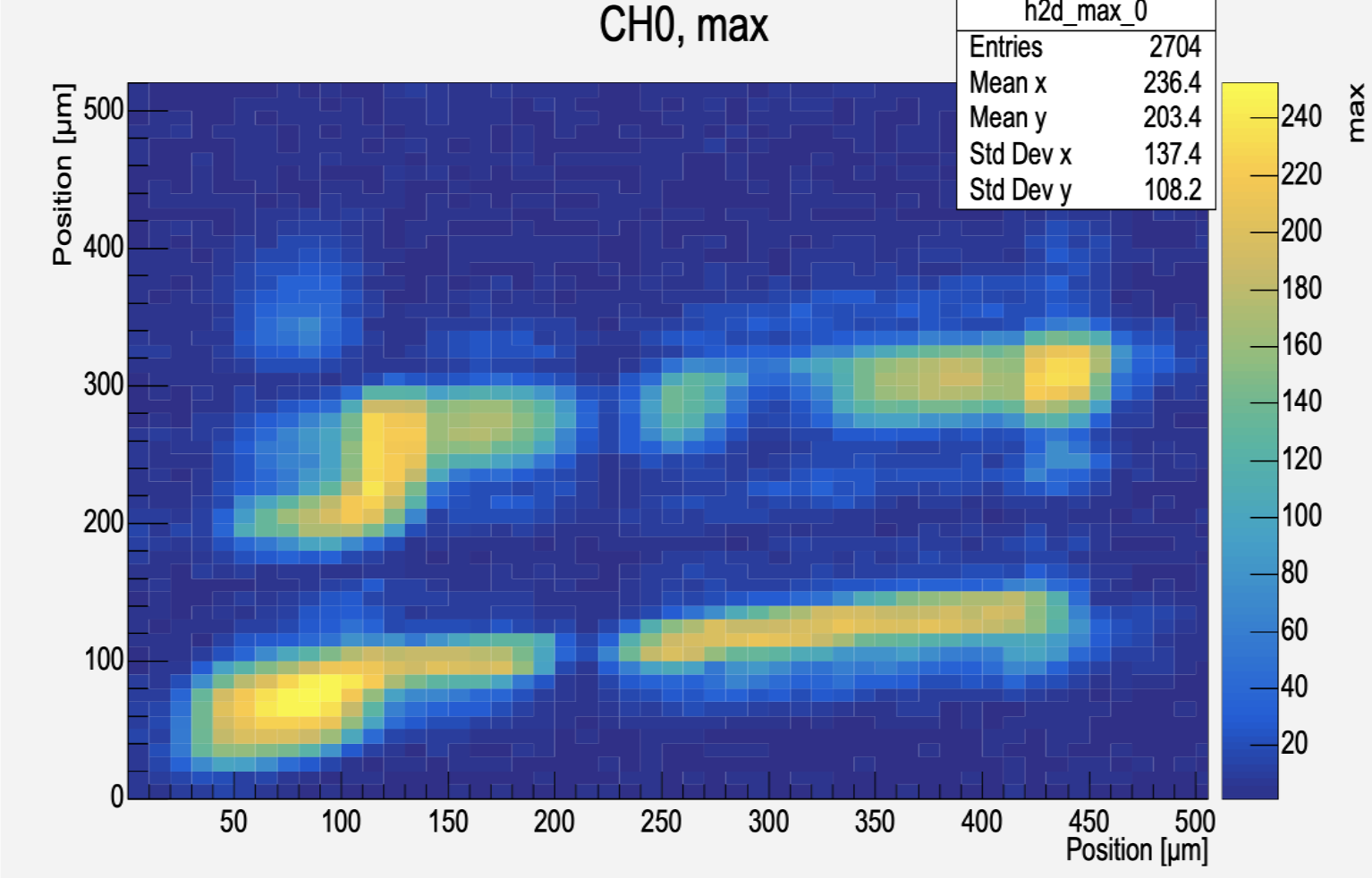}
   \end{tabular}
   \end{center}
   \caption{ \label{fig:front} 
        A medium-resistivity ($300\pm100~\Omega*$cm) chip as scanned with an infrared laser on the implant surface shows clear structure reminiscent of the chip's metalization layers. a) The design metalization has clear periodic structure which is repeated for every individual pixel. The top-most layer is shown in dark red and the lowest layer is shown in bright red. The implanted n-well can be seen in purple between other structures. b) Analog data as collected from AstroPix while an infrared laser scanned across the implanted surface of the chip in 10~$\mu$m steps. The color bar indicates the height of the measured analog pulse, where lighter yellow colors are taller peaks (correlated to higher measured energy deposits). The origin is arbitrary. The chip metalization structure is clearly seen in the recorded analog data.}
\end{figure}

However, the medium-resistivity chip was used to imagine the chip surface (see Fig.~\ref{fig:front}). The structure of the top metalization layer can clearly be seen with the pixel's analog output data.

The most reliable eTCT data was collected from a low-resistivity chip ($25\pm8.25~\Omega*$cm) where depletion is not expected to reach the desired $500~\mu$m (see Fig.~\ref{fig:depl_theory}). Bias voltages of - (40-230)~V were considered. The $z$ dimension of the laser was defined by preliminary scans in $xz$. The $z$ location that maximized AstroPix analog output was chosen for further edge testing in the $xy$ plane and held consistent across all subsequent runs.

\begin{figure} [ht]
   \begin{center}
   \begin{tabular}{c} %% tabular useful for creating an array of images 
   \includegraphics[height=8cm]{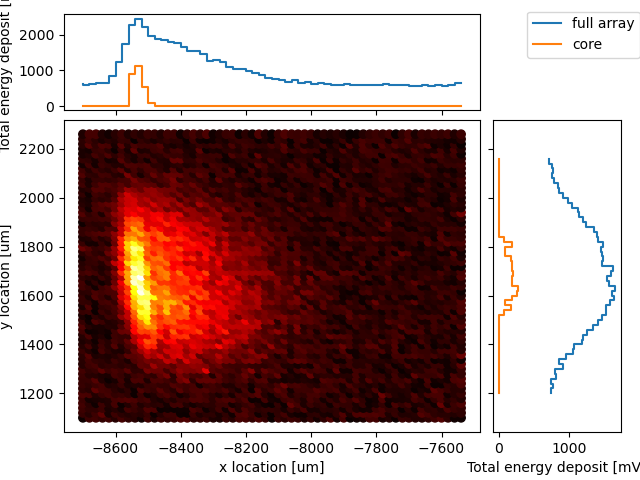}
   \end{tabular}
   \end{center}
   \caption{ \label{fig:depl_meas} 
        The analog response of the edge AstroPix pixel differs in height as the laser location (in Fig.~\ref{fig:etct}) is moved. Scanning through the depth of the wafer (increasing $x$) shows how quickly depletion falls off. Scanning across the pixel (increasing $y$) indicates charge sharing between pixels. The profile histograms on the top and right show the average analog response in that $x$ or $y$ location for the full scan area (in blue) as well as the 'core' scan area (in orange), as defined in the text. This figure shows the low-resistivity chip response when biased to -180V.}
\end{figure}

Using preliminary runs to define an appropriate origin, scans in $xy$ were duplicated over a range of bias voltages. After each scan, the chip was returned to the same origin position before the next scan. One measurement, the average of 100 waveforms from only the pixel on the edge of the array that the laser directly interacted with, was made every 20~$\mu$m and the average pulse height recorded (see Fig.~\ref{fig:depl_meas} where brighter yellow colors indicate larger pulse amplitudes than darker red colors). Profile histograms are then created from this scan data by considering the average pulse height recorded in each row or column and using that average value as one histogram bin. The profile histogram in the top of Fig.~\ref{fig:depl_meas} shows a profile in $x$ where the $y$ scan points are averaged over, and the inverse is true for the $y$ profile on the right.

The silicon wafer is 720~$\mu$m thick, and the data of Fig.~\ref{fig:depl_meas} illustrates this. The distance from the surface of the chip, around $x=-8600$, to the absence of pulse heights over the noise level around $x=-7900$ is the expected $720~\mu$m. The $y$ extent is larger than expected with the $500~\mu$m pixel pitch. This could be due to charge sharing between pixels and relative angles within the eTCT measurement system resulting in a larger 'effective' pixel pitch.

The potential for charge sharing, as indicated by analog hits that extend beyond the expected pixel pitch, motivate a stricter requirement for continued analysis of the eTCT scan data. Rather than considering $x$ profile histograms from the full scan to understand the growth of the depletion region, stricter conditions are met. First, a noise floor is defined as the average of the three right-most $x$ columns. Any scan point with pulse amplitude less than this noise floor is disregarded. Then, the single brightest scan point (with the largest pulse amplitude) is used to define the 'core' deposit region. The measurement of a scan point is included as part of the 'core' provided that is both larger than the noise floor and at least $75\%$ as bright as the brightest point. This three-quarters condition is only meant to guide this continued analysis, and its absolute value will be further optimized in future studies. Another profile histogram is created considering only scan points that pass the criteria described above to fall in the 'core' region. This 'core' histogram is shown in orange in Fig.~\ref{fig:depl_meas} where the scaling from noise is clearly eliminated and a clearer return to baseline can be seen. With this 'core' definition, the $y$ extent of the pixel does agree more closely with the expected $500~\mu$m.

\begin{figure} [ht]
   \begin{center}
   \begin{tabular}{ccc} %% tabular useful for creating an array of images 
   \includegraphics[height=6cm]{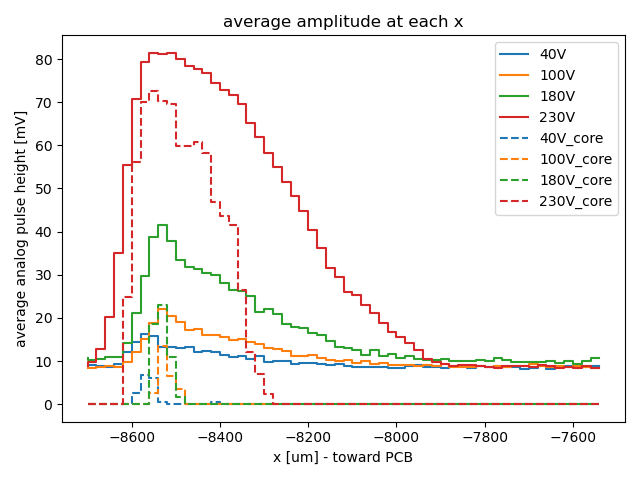}&& b) \includegraphics[height=4.75cm]{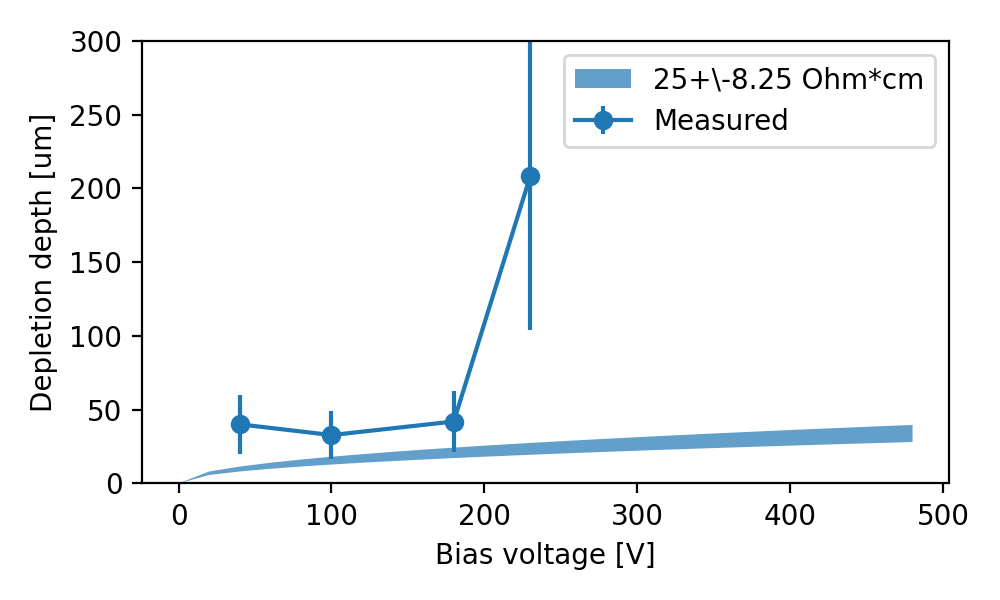}
   \end{tabular}
   \end{center}
   \caption{ \label{fig:depl_meas2} 
        The measured AstroPix response is impacted by applied bias voltage shows higher depletion than expected from theory. a) At each $x$ laser scan location, every $y$ measurement (as in Fig~\ref{fig:depl_meas}) is averaged. A 'core' region is seeded with the brightest scan point and any point measuring a pulse height of at least 75\% of the seed is considered in the 'core' average. b) Each 'core' histogram of a) is fit with a Gaussian and the full-width half max (FWHM) is shown compared to the theoretical depletion. 50\% error bars are added to data points. Measurements imply larger depletion depths than expected from theory, likely resulting from charge sharing between pixels and additional sources of error in the experimental setup. }
\end{figure}

The full-scan average and 'core' $x$ profiles for multiple different bias voltages are shown in Fig.~\ref{fig:depl_meas2} a). The $x$ profile of a fully depleting chip is expected to look like a step function that sharply cuts off when the laser is scanned into an area that is not within the depletion region. The AstroPix response rises sharply, as expected, but a gradual decline in $x$ rather than a sharp step function. Lower full-scan average and 'core' average values are seen with lower bias voltages, as expected. 

A preliminary depletion depth value is extracted from the 'core' average $x$ profiles of Fig.~\ref{fig:depl_meas2} by fitting a Gaussian distribution to the histogram, extracting the standard deviation fit parameter $\sigma$, and converting to a full-width half maximum (FWHM) as FWHM$=2.355\sigma$. These extracted depletion values are shown in Fig.~\ref{fig:depl_meas2} b) as compared to theoretical calculations from Fig.~\ref{fig:depl_theory}. $50\%$ error bars are shown, as a full error analysis is still in progress. The measured values still appear larger than the expected theory. This could be due to some combination of the following effects that have not yet fully been taken into account: charge sharing, relative angles in the measurement setup, reflections in the measurement apparatus, 'core' region definition.

These results should be considered as a first, preliminary consideration of AstroPix depletion and are pending a final  analysis with more thorough considerations of error.

%%%%%%%%%%%%%%%%%%%%%%%%%%%%%%%%%%%%%%%%%%%%%%%%%%%%%%%%%%%%%%%%%%%%%%%%%%%%%%%%%%%%%%%%%%%
\subsection{Next Steps}
\label{ssec:etct_steps}

The many unaccounted for uncertainties in this eTCT measurement justify repeating this measurement with more chips in the future. An updated stage and laser setup is already in place at SCIPP which will provide more information about relative stage location between runs. Fewer reflections will be present in the system after the design of a custom PCB explicitly to be used for this measurement which moves the AstroPix array edge to the edge of the PCB and minimizes reflective components on the PCB surface. The edge of the AstroPix chip incident with the laser will also be polished before the measurement, as the technique for polishing has been tested and documented at SCIPP. The high-resistivity chips should also be tested again with the updated understanding of PCB mounting, as discussed in Sec.~\ref{sec:highres}.

Finally, TCAD simulations of the AstroPix pixel and its expected depletion depth are underway and should be used for data comparison in addition to the theoretical curve of Eq.~\ref{eq:depl}.

Future measurements can also be made with future versions of AstroPix, including AstroPix\_v4\cite{striebig_v4} which is currently in-hand and awaiting testing and AstroPix\_v5 which is scheduled to be fabricated in 2024. 

%%%%%%%%%%%%%%%%%%%%%%%%%%%%%%%%%%%%%%%%%%%%%%%%%%%%%%%%%%%%%%%%%%%%%%%%%%%%%%%%%%%%%%%%%%%%%%%%%%%%%%%%%%%%%%%%%%%%%%%%%%%%%%%%%%%%%%%%%%%%%%%%%%%%%%%%%%%%%%%%%%%%%%%%%%%%%%%%%%%%%%%%%%%%%%%%%%%%%%%%%%%%%%%%%%%%%%%%%%%%%%%%%%%%%%%%%%%%%%%%%%%%%%%%%%%%%%%%%%%%%%%%%%%%%%%%%
\section{CONCLUSIONS AND OUTLOOK}
\label{sec:concl}

The AstroPix HVCMOS pixelated silicon chip is designed to deplete a depth of 500~$\mu$m in order to achieve a dynamic range of 25-700 keV per pixel. The current version under test, AstroPix\_v3, does not yet meet these requirements however two studies summarized in this work illustrate the work done to understand this problem and inform future versions. 

High leakage current off the high-voltage bias line of chips fabricated on high-resistivity wafers was found to be due, in part, to the mounting of chips onto readout PCBs. Inherent backside voltage gradients are exaggerated when the chips are mounted with conductive glue or tape. Utilizing an insulating material between the chip and PCB enables higher bias voltages to be applied to the chips before reaching high currents (1~$\mu$A). Testing of the resulting data quality is still underway.

A direct measurement of depletion was also started using edge Transition Current Technique of scanning a focused infrared laser down the side of the chip from the implant surface to the backside. This test was done with chips of multiple wafer resistivities, but experimental setup limitations and conducting mounting tape for the high-resistivity chip left only the lowest-resistivity chips with the lowest expected depletion depths viable for further analysis. Preliminary analysis shows the known $y$ extent of the pixels but indicates a potential for charge sharing between pixels. Further analysis, including a robust consideration of sources of uncertainty, is still underway.

%%%%%%%%%%%%%%%%%%%%%%%%%%%%%%%%%%%%%%%%%%%%%%%%%%%%%%%%%%%%%%%%%%%%%%%%%%%%%%%%%%%%%%%%%%%%%%%%%%%%%%%%%%%%%%%%%%%%%%%%%%%%%%%%%%%%%%%%%%%%%%%%%%%%%%%%%%%%%%%%%%%%%%%%%%%%%%%%%%%%%%%%%%%%%%%%%%%%%%%%%%%%%%%%%%%%%%%%%%%%%%%%%%%%%%%%%%%%%%%%%%%%%%%%%%%%%%%%%%%%%%%%%%%%%%%%%
%\appendix    %>>>> this command starts appendixes

%\section{APPENDIX SECTION}

%%%%%%%%%%%%%%%%%%%%%%%%%%%%%%%%%%%%%%%%%%%%%%%%%%%%%%%%%%%%%%%%%%%%%%%%%%%%%%%%%%%%%%%%%%%%%%%%%%%%%%%%%%%%%%%%%%%%%%%%%%%%%%%%%%%%%%%%%%%%%%%%%%%%%%%%%%%%%%%%%%%%%%%%%%%%%%%%%%%%%%%%%%%%%%%%%%%%%%%%%%%%%%%%%%%%%%%%%%%%%%%%%%%%%%%%%%%%%%%%%%%%%%%%%%%%%%%%%%%%%%%%%%%%%%%%%
\acknowledgments % equivalent to \section*{ACKNOWLEDGMENTS}       
 
This work is funded in part by 18-APRA18-0084 and 
20-RTF20-0003. AS acknowledges that research was sponsored by NASA through a contract with ORAU.

% References
\bibliography{astropix} % bibliography data in astropix.bib
\bibliographystyle{spiebib} % makes bibtex use spiebib.bst

\end{document}